\documentclass[10pt,twocolumn,amssymb,amsmath,aps,pra]{revtex4-1}
\usepackage{graphicx}

\begin{document}

\title{Comment on ``Analysis of recent interpretations of the Abraham-Minkowski problem"}
\date{July 16, 2019}
\author{Mikko Partanen}
\author{Jukka Tulkki}
\affiliation{Engineered Nanosystems Group, School of Science, Aalto University, P.O. Box 12200, 00076 Aalto, Finland}

\begin{abstract}
In a recent paper [I.~Brevik, Phys.~Rev.~A \textbf{98}, 043847 (2018)], Brevik analyzed the experiment by Kundu \emph{et al.} [A.~Kundu \emph{et al.}, Sci.~Rep.~\textbf{7}, 42538 (2017)] reporting deformation of a graphene oxide (GO) film after it has been irradiated by a laser beam. The two-dimensional atomic force microscope (AFM) line scanning of the deformation of the GO film after switching off the laser beam takes by far too much time for any elastic changes to remain in the AFM scans. Thus, the changes in the GO film are irreversible and the optoelastic model used by Brevik is not applicable. The rough estimates of the kinetic energy and displacement of atoms by the optical force of a light pulse calculated by Brevik are correct, but in making a comparison with the corresponding high-precision results for the kinetic energy and displacement of atoms in our work [M.~Partanen \emph{et al.}, Phys.~Rev.~A \textbf{95}, 063850 (2017)], the kinetic energy of atoms is confused with their rest energy. The atoms and their masses are displaced forward by the field and their displaced rest energies give rise to an energy flux. The difference of arguments between ours and Brevik's culminates on the question whether the flux of rest energy caused by the displacement of the medium moving with light should be included in the total energy flux. We also show that the four-divergence of the stress-energy-momentum tensor of the mass polariton theory of light is zero.
\end{abstract}

\maketitle


\section{Introduction}

In a recent paper \cite{Brevik2018b}, Brevik analyzed an interesting experiment by Kundu \emph{et al.}~\cite{Kundu2017}, who reported the deformation of a graphene oxide (GO) film after irradiating it with a laser beam. Kundu \emph{et al.}~concluded that the observed deformation supports the Abraham model of the momentum of light. Brevik also made some critical observations regarding the mass polariton (MP) theory of light presented by Partanen \emph{et al.}~\cite{Partanen2017c}. We start by considering these comments and then, at the end of our paper, we briefly remark on the experiment of Kundu \emph{et al.}~and in particular its relation to the model that Brevik used to calculate the bending of the GO film under the influence of a laser beam.


\section{Calculating the kinetic energy of atoms}

On p.~1 of his paper \cite{Brevik2018b}, Brevik claimed that, in our work \cite{Partanen2017c}, we would have predicted ``transfer of quite a large mechanical energy from the pulse to the medium, of the same order of magnitude as the field energy itself." This claim is unsound. In Ref.~\cite{Partanen2017c}, or in follow-up works \cite{Partanen2017e,Partanen2018a,Partanen2018b}, we have nowhere claimed that a light wave would be associated with a mechanical energy, kinetic energy, or elastic energy, of the same order of magnitude as the field energy.

On the contrary, on p.~12 of our work \cite{Partanen2017c}, we calculated, for a light pulse having a total field energy of 5 mJ, kinetic energy of the medium atoms to be $3.6\times 10^{-16}$ eV per photon, which makes $1.2\times 10^{-15}$ mJ for the whole light pulse. Thus, our work \cite{Partanen2017c} proved that the kinetic energy is nonzero, but a vanishingly small part of the field energy. This follows directly from the classical expression of the kinetic energy of atoms in Eq.~(25) of Ref.~[3] used in the optoelastic continuum dynamics (OCD) of light. The OCD model is based on combining the electrodynamics of continuous media, elasticity theory, and Newtonian mechanics. It can be used to calculate the kinetic energy of the medium for an arbitrary light pulse. However, for all technologically realizable light pulses and highly transparent materials, the kinetic energy of atoms remains extremely small in relation to the field energy. For the OCD simulations, see also the section discussing how atoms with negligible kinetic energy can carry a mass with the velocity of light.

In our work \cite{Partanen2017c} we have also shown that the kinetic energy of the mass density wave (MDW) part of the MP is negligibly small in the case of a single photon (see p.~3 and 4 of Ref.~\cite{Partanen2017c}). Since the momentum of the medium part of the MP can at most be of the same order of magnitude as the momentum of the photon in vacuum, the kinetic energy is $E_\mathrm{k}\le(\hbar\omega/c)^2/2m_0$, where $m_0$ is the mass of the smallest structural unit in the medium. For example, for $\hbar\omega=0.80$ eV ($\lambda_0=1550$ nm) and $m_0=4.7\times 10^{-26}$ kg, the mass of a single silicon atom, we obtain $E_\mathrm{k}\le 1.2\times 10^{-11}$ eV. The smallness of the kinetic energy of the medium part of the MP is a direct consequence of the extremely small momentum to energy ratio of a photon as compared to the momentum to kinetic energy ratio of a particle having a rest mass. Thus, in our work \cite{Partanen2017c} we have shown using two independent approaches that the kinetic energy of atoms in the MDW is a vanishingly small part of the field energy.

\section{\label{sec:gedanken}Theoretical gedanken experiment}

In Eq.~(35) of his work \cite{Brevik2018b}, Brevik found for an average momentum of atoms under the influence of the optical force of a light pulse having duration $\tau$, an expression
\begin{equation}
 \Delta p=\frac{n^2-1}{Nc^2}(\mathbf{E}\times\mathbf{H})_x,
\label{eq:atommomentum}
\end{equation}
where $N$ is the number density of particles, $n$ is the refractive index, and $\mathbf{E}\times\mathbf{H}$ is the Poynting vector, in which $\mathbf{E}$ and $\mathbf{H}$ are the electric and magnetic fields, respectively. The momentum density obtained by multiplying the momentum of a single atom in Eq.~\eqref{eq:atommomentum} by the number density $N$ agrees with the momentum density of the MDW in the integrand of Eq.~(13) of our work \cite{Partanen2017c}.  Brevik arrived at the same conclusion as we did in our work \cite{Partanen2017c} that the kinetic energy of atoms is extremely small.

The fundamental property of the MP theory of light presented in Ref.~\cite{Partanen2017c} is that the optical force gives rise to the time- and position-dependent displacement of the medium atoms (see, e.g., Figs.~5 and 7 of Ref.~\cite{Partanen2017c} and Figs.~4 and 5 of Ref.~\cite{Partanen2017e}). Brevik also reproduced these results for his schematic light pulse and obtained the correct average total displacement of an atom in the medium upon a pulse of duration $\tau$, given by $l=(\Delta p/m_0)\tau=v_\mathrm{a}\tau$, where $m_0$ is the rest mass of an atom and $v_\mathrm{a}$ is the atomic velocity.

Would Brevik have continued his theoretical gedanken experiment a step further, he would have found what we guess he means by the ``mechanical" energy of the same order of magnitude as the field energy. The total displaced volume of atoms in the case of his example pulse is straightforwardly given by $V=Al$, where $A$ is the cross-sectional area of the pulse. Therefore, following Brevik's example, the effective displaced medium mass becomes
\begin{equation}
 \delta M=\rho V=\frac{n^2-1}{c^2}(\mathbf{E}\times\mathbf{H})_x A\tau=\frac{n^2-1}{c^2}E_\mathrm{field},
\label{eq:mass}
\end{equation}
 where we have used the conventional relations $Nm=\rho$ and $(\mathbf{E}\times\mathbf{H})_x A\tau=E_\mathrm{field}$, in which $E_\mathrm{field}$ is the field energy. This mass of the shifted density of atoms moving with the light pulse is, in the special theory of relativity (STR), equivalent to energy $\delta Mc^2=(n^2-1)E_\mathrm{field}$, which must be added to the energy flux of the field to obtain the total energy flux of the light pulse including all possible forms of energy.

\begin{figure}
\centering
\includegraphics[width=0.85\columnwidth]{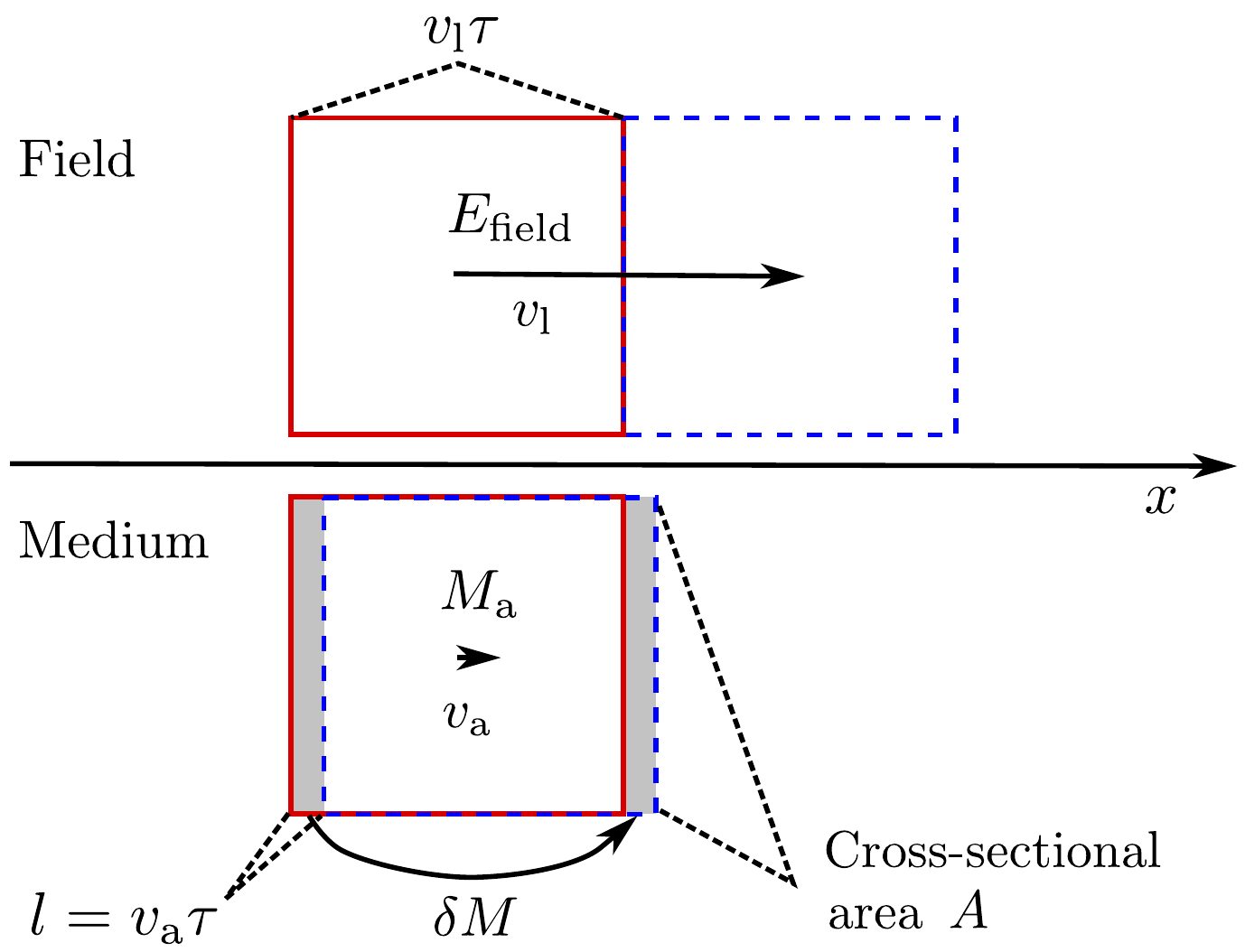}
\caption{\label{fig:masstransfer}
Illustration of the mass transfer effect and the related different scales of mass shifts. The red solid blocks denote the energy $E_\mathrm{field}$ of the field (top) and the mass $M_\mathrm{a}=\rho Av_\mathrm{l}\tau$ of atoms (bottom) in the volume of a light pulse at $t=0$. The light pulse is propagating to the right. The blue dashed blocks denote the displaced positions of $E_\mathrm{field}$ and $M_\mathrm{a}$ at $t=\tau$. The mass $M_\mathrm{a}$ is moved due to the optical force. In time $\tau$, the field energy has moved a distance $v_\mathrm{l}\tau$ with velocity $v_\mathrm{l}=c/n$. In this same time, the center of mass $M_\mathrm{a}$ has moved a distance $l=v_\mathrm{a}\tau$ with small atomic velocity $v_\mathrm{a}$. However, if we compare the initial and final distributions of matter to each other, we observe that a small effective mass $\delta M=M_\mathrm{a}v_\mathrm{a}/v_\mathrm{l}$ (and the related rest energy $\delta Mc^2$, not shown) in the left gray area has moved forward a distance $v_\mathrm{l}\tau\gg l$ with light. Therefore, the effective mass transfer with light is an unavoidable consequence of the optical force. The results of the detailed numerical OCD simulations are in full agreement with this simple schematic picture.}
\end{figure}

In the STR, an atom moving under the influence of the optical force does not only carry its momentum and kinetic energy. When an atom is moving, its mass is also shifted and so is the rest energy related to this mass. Thus, in time $\tau$, a single atom gives rise to the shift of energy $m_0c^2$ plus the negligible kinetic energy by a length $l$. In Fig.~\ref{fig:masstransfer} we illustrate how counting together the very small shift of all atoms in the volume of the light pulse in time $\tau$ gives rise to an effective shift of mass equal to $\delta M=\rho V$ by $v_\mathrm{l}\tau\gg l$ along the $x$ axis. In the STR, the shift of $\delta M$ corresponds to shift of energy $\delta Mc^2$ by $v_\mathrm{l}\tau$. Thus, one can conclude that the effective energy density of moving atoms is equal to $\delta Mc^2/V=(n^2-1)(\mathbf{E}\times\mathbf{H})_x\tau/l$ plus the negligible kinetic energy of all atoms. This energy density must be added to the energy density of the field to obtain the total energy density of the light pulse. In contrast, in his theoretical gedanken experiment, Brevik concluded that, since the kinetic energy of atoms moving with the light pulse is negligibly small, the total energy density of light is equal to the energy density of the field alone.

\section{\label{sec:ocd}How atoms with negligible kinetic energy can carry a mass with the velocity of light}

The fundamental starting point of the OCD model introduced by us in Ref.~\cite{Partanen2017c} is that we write the equation of motion for the medium under the influence of the optical forces associated with the light pulse and the elastic forces that also become effective after the optical force has displaced the atoms from their equilibrium positions. Thus, by solving this dynamical equation, we automatically obtain the displacement and velocity fields of the medium elements for an arbitrary light pulse. The OCD model also reproduces the elastic waves following from the impact of a light pulse \cite{Pozar2018}.

The atomic displacements in the OCD simulation also lead to the perturbation of the mass density of the medium. The excess mass density obtained by solving Newton's equation of motion is given by \cite{Partanen2017c}
\begin{equation}
 \rho_\mathrm{MDW}=\rho_\mathrm{a}-\rho_0,
\label{eq:densityequality}
\end{equation}
where $\rho_\mathrm{a}$ is the true perturbed mass density of atoms and $\rho_0$ is the atomic mass density in the absence of the light pulse. One can show analytically in the case of a plane wave or by computer simulations using the OCD theory that the volume integral of the excess mass density of atoms in Eq.~\eqref{eq:densityequality} over the light pulse is equal to $\delta M$ given in Eq.~\eqref{eq:mass}, both in the case of the gedanken experiment by Brevik and in the case of a general light pulse. 

Using the standard formula for the momentum density of atoms in the laboratory frame, our simulations give, within the numerical accuracy of the computations, typically seven to eight digits, the equality \cite{Partanen2017c}
\begin{equation}
 \rho_\mathrm{a}\mathbf{v}_\mathrm{a}=\rho_\mathrm{MDW}\mathbf{v}_\mathrm{l},
\label{eq:momentumequality}
\end{equation}
where $\mathbf{v}_\mathrm{a}$ and $\mathbf{v}_\mathrm{l}$ are the atomic and light velocity vectors. On the left-hand side of Eq.~\eqref{eq:momentumequality} we have the large true atomic density and small atomic velocity, while on the right-hand side we have the small excess mass density of atoms, and the large velocity of light in the medium. Therefore, the equality in Eq.~\eqref{eq:momentumequality} essentially explains how the small collective motion of atoms forms a MDW that propagates with the velocity of light in the medium and carries with itself the excess atomic density of Eq.~\eqref{eq:densityequality} corresponding to the effective mass $\delta M$ in Eq.~\eqref{eq:mass} and also the related rest energy $\delta Mc^2$. This is in full agreement with the schematic illustration in Fig.~\ref{fig:masstransfer}.

In summary, we can write the total energy density $W_\mathrm{MP}=W_\mathrm{field}+W_\mathrm{MDW}$ and the total momentum density $\mathbf{G}_\mathrm{MP}=\mathbf{G}_\mathrm{field}+\mathbf{G}_\mathrm{MDW}$ of the MP theory as sums of the  field and the MDW contributions, given by \cite{Partanen2017c} 
\begin{equation}
 W_\mathrm{field}=\frac{1}{2}(\mathbf{E}\cdot\mathbf{D}\!+\!\mathbf{H}\cdot\mathbf{B}),\hspace{0.5cm}
W_\mathrm{MDW}=\rho_\mathrm{MDW}c^2,
\label{eq:mpenergy}
\end{equation}
\begin{equation}
 \mathbf{G}_\mathrm{field}=\frac{\mathbf{E}\times\mathbf{H}}{c^2},\hspace{0.5cm}
\mathbf{G}_\mathrm{MDW}=\rho_\mathrm{MDW}\mathbf{v}_\mathrm{l},
\label{eq:mpmomentum}
\end{equation}
where $\mathbf{D}$ and $\mathbf{B}$ are the electric and magnetic flux densities, respectively. In the discussion of the stress-energy-momentum (SEM) tensor of the MP theory, we will also use the following useful relations, which apply with high accuracy in the laboratory frame [see Eqs.~(13) and (17) of Ref.~\cite{Partanen2017c}]:
\begin{equation}
 W_\mathrm{MDW}^\mathrm{(L)}=\frac{n^2-1}{2}(\mathbf{E}\cdot\mathbf{D}\!+\!\mathbf{H}\cdot\mathbf{B}),
\label{eq:mdwenergy}
\end{equation}
\begin{equation}
 \mathbf{G}_\mathrm{MDW}^\mathrm{(L)}=\frac{n^2-1}{c^2}\mathbf{E}\times\mathbf{H}.
\label{eq:mdwmomentum}
\end{equation}
Note that the integral of the rest energy density of the atomic MDW in Eq.~\eqref{eq:mdwenergy} is equal to $\delta Mc^2$, with $\delta M$ given in Eq.~\eqref{eq:mass}. Thus, Eq.~\eqref{eq:mdwenergy} is in agreement with Brevik's analysis. The momentum density of the MDW in Eq.~\eqref{eq:mdwmomentum} is also consistent with Brevik's analysis as it is equal to Eq.~(34) of Ref.~\cite{Brevik2018b}.

\section{\label{sec:mptensor}Lorentz covariance and the MP SEM tensor}

In his work \cite{Brevik2018b} on p.~4, Brevik also criticized our work \cite{Partanen2017c} by stating that, due to the term $\delta Mc^2$, the ``values of $E_\mathrm{MP}$ and $p_\mathrm{MP}$ do not allow one to use the Lorentz transformation, as they are not the energy and momentum components of an energy-momentum tensor whose four-divergence is zero." However, Brevik provided no arguments to support his claim that the four-divergence of the SEM tensor of the MP would be nonzero.

We have introduced the SEM tensor of the MP theory concisely in Appendix B of our work \cite{Partanen2017c}. In this analysis, we neglect the negligibly small kinetic and elastic energies of atoms in the laboratory frame. Thus, in the MDW, we include only the dominant terms, which are of the first order in the small atomic velocity $v_\mathrm{a}$. These terms are the MDW rest energy density and the MDW momentum density that are moving with light. Following Ref.~\cite{Partanen2017c}, the total MP SEM tensor in the laboratory frame is a sum of the SEM tensor $\mathbf{T}_\mathrm{field}$ of the field and the SEM tensor $\mathbf{T}_\mathrm{MDW}$ of the atomic MDW, given by
 \begin{equation}
  \mathbf{T}_\mathrm{MP}=\mathbf{T}_\mathrm{field}+\mathbf{T}_\mathrm{MDW}.
  \label{eq:tensorsum}
 \end{equation}

Including the contributions of both the field and the atomic MDW and using Eqs.~\eqref{eq:mpenergy}--\eqref{eq:mdwmomentum}, the total energy-momentum tensor of the field and the MDW is given in the laboratory frame by \cite{Partanen2017c}
\begin{align}
  &\mathbf{T}_\mathrm{MP}\nonumber\\
  &=\bigg[\begin{array}{cc}
   \frac{1}{2}(\mathbf{E}\!\cdot\!\mathbf{D}\!+\!\mathbf{H}\!\cdot\!\mathbf{B})+\rho_\mathrm{MDW}c^2 & \frac{1}{c}(\mathbf{E}\times\mathbf{H})^\mathrm{T}+\rho_\mathrm{MDW}\mathbf{v}_\mathrm{l}^\mathrm{T}c\\
   \frac{1}{c}\mathbf{E}\times\mathbf{H}+\rho_\mathrm{MDW}\mathbf{v}_\mathrm{l}c & -\boldsymbol{\sigma}
  \end{array}\bigg]\nonumber\\
  &=\bigg[\begin{array}{cc}
   \frac{n^2}{2}(\mathbf{E}\cdot\mathbf{D}\!+\!\mathbf{H}\cdot\mathbf{B}) & \frac{n^2}{c}(\mathbf{E}\times\mathbf{H})^\mathrm{T}\\
   \frac{n^2}{c}\mathbf{E}\times\mathbf{H} & -\boldsymbol{\sigma}
  \end{array}\bigg],
  \label{eq:tensormp}
\end{align}
where the superscript $\mathrm{T}$ indicates the transpose and $\boldsymbol{\sigma}$ is the electromagnetic stress tensor, given by a $3\times 3$ matrix as \cite{Jackson1999}
 \begin{equation}
  \boldsymbol{\sigma}=\mathbf{E}\otimes\mathbf{D}+\mathbf{H}\otimes\mathbf{B}
  -\frac{1}{2}(\mathbf{E}\cdot\mathbf{D}\!+\!\mathbf{H}\cdot\mathbf{B})\mathbf{I}.
 \end{equation}
 Here $\otimes$ denotes the outer product and $\mathbf{I}$ is the $3\times3$ unit matrix. In the last row of Eq.~\eqref{eq:tensormp}, we have expressed the MDW quantities in terms of the field quantities using Eqs.~\eqref{eq:mdwenergy} and \eqref{eq:mdwmomentum}.

The proof of the zero four-divergence of the MP SEM tensor in Eq.~\eqref{eq:tensormp} is straightforward in the general case by using Maxwell's equations and the well-known vector calculus identities.  This proof is presented in the Appendix. We conclude that the four-divergence of the MP SEM tensor is zero, in contrast to what Brevik claimed.

In the works cited by Brevik \cite{Gordon1923,Leonhardt2006b,Leonhardt2010}, the well-known angular momentum conservation law problem of the conventional asymmetric Minkowski SEM tensor has been artificially solved by introducing the Gordon metric, which depends on the permittivity and permeability of materials, and consequently, leads to artificial gravitational fields that are not physically true in the sense of the general theory of relativity. Therefore, these works do not solve the angular momentum conservation law problem of the Minkowski SEM tensor in the physical space-time whose metric is only modified by the curvature of the space-time due to true gravitational fields. For a concise summary of the MP SEM tensor and the Minkowski SEM tensor, we refer to Table I of Ref.~\cite{Partanen2019a}.

\section{Kundu \emph{et al.}~experiment}

In the Kundu \emph{et al.}~experiment, the authors carried out atomic force microscopy (AFM) measurements of the GO film and in particular reported the ``AFM height image of GO surface after focused laser irradiation at different spots for various laser
power." See Fig.~3 of Ref.~\cite{Kundu2017} for the AFM images, which are taken well after the laser irradiation. If the changes were elastic, the bending of the surface would not be visible after the laser irradiation when the two-dimensional AFM line scanning of the images, requiring mechanical shifting the AFM tip, was carried out. We conclude that the elasticity theory used by Brevik \cite{Brevik2018b} is not applicable in the analysis of the irreversible changes observed by Kundu \emph{et al.}~\cite{Kundu2017}. A possible explanation of the deformed GO surface is provided by the low density of the GO film, given by $\rho=750$ kg/m$^3$ \cite{Rioboo2015}, which suggests a complex material structure, which may have a very low irradiation damage threshold. The irreversibility of the deformation of the GO film also means that the interaction of light with the medium leads to changes in the molecular structure of the material. Therefore, the optical force concepts behind the Abraham and Minkowski momenta are not enough in the analysis of this experiment. Thus, we agree with Brevik, that the Kundu \emph{et al.}~experiment is unlikely to provide the experimental resolution of the Abraham-Minkowski paradox of light.

\section{Conclusion}

Brevik has overlooked that the smallness of the kinetic energy of atoms was already calculated and well documented in our original work \cite{Partanen2017c}. The remaining difference between the theories of Brevik and ours is that we argue on the basis of the fundamental principles of the special theory of relativity and relativistic mechanics that the rest energy of the moving medium elements must be included in the total energy flux of light. In Brevik's model, this mass energy flux is excluded without giving detailed arguments that would justify it. We have also shown in the Appendix that the four-divergence of the MP SEM tensor is zero. For detailed classical field-theoretical foundations of the MP theory of light, see the follow-up works \cite{Partanen2019a,Partanen2019b}. We have finally argued that the GO surface deformations observed in the experiment of Kundu \emph{et al.}~\cite{Kundu2017} are irreversible. Thus, the deformation cannot be analyzed by elasticity theory and conventional optical forces concepts, which do not account for breaking
of chemical bonds, as done in the work of Brevik.

\appendix

\section{Four-divergence of the MP SEM tensor}

Here we present a proof of the zero four-divergence of the MP SEM tensor in the laboratory frame of a nondispersive lossless medium. We use the well-known constitutive relations $\mathbf{D}=\varepsilon\mathbf{E}$ and $\mathbf{B}=\mu\mathbf{H}$, in which $\varepsilon$ and $\mu$ are the permittivity and permeability of the medium that are assumed to be constant in a homogeneous isotropic medium. Then, the four-divergence of the first row vector of the MP SEM tensor in the laboratory frame in Eq.~\eqref{eq:tensormp} can be written as
\begin{align}
 &\frac{1}{c}\frac{\partial}{\partial t}\Big[\frac{n^2}{2}(\varepsilon\mathbf{E}^2+\mu\mathbf{H}^2)\Big]
+\nabla\cdot\Big[\frac{n^2}{c}\mathbf{E}\times\mathbf{H}\Big]\nonumber\\
&=\frac{n^2}{c}\Big[\frac{\varepsilon}{2}\frac{\partial}{\partial t}\mathbf{E}^2+\frac{\mu}{2}\frac{\partial}{\partial t}\mathbf{H}^2
+(\nabla\times\mathbf{E})\cdot\mathbf{H}-\mathbf{E}\cdot(\nabla\times\mathbf{H})\Big]\nonumber\\
&=\frac{n^2}{c}\Big[\frac{\varepsilon}{2}\frac{\partial}{\partial t}\mathbf{E}^2+\frac{\mu}{2}\frac{\partial}{\partial t}\mathbf{H}^2
-\mu\mathbf{H}\cdot\frac{\partial}{\partial t}\mathbf{H}-\varepsilon\mathbf{E}\cdot\frac{\partial}{\partial t}\mathbf{E}\Big]\nonumber\\
&=\frac{n^2}{c}\Big[\frac{\varepsilon}{2}\frac{\partial}{\partial t}\mathbf{E}^2+\frac{\mu}{2}\frac{\partial}{\partial t}\mathbf{H}^2
-\frac{\mu}{2}\frac{\partial}{\partial t}\mathbf{H}^2-\frac{\varepsilon}{2}\frac{\partial}{\partial t}\mathbf{E}^2\Big]\nonumber\\
&=0.
\label{eq:apx1}
\end{align}
In the first equality, we have taken the constant factors outside the derivatives and applied the vector differential identity for the divergence of the cross product in the second term. In the second equality, we have used Faraday's law $\nabla\times\mathbf{E}=-\frac{\partial}{\partial t}\mathbf{B}=-\mu\frac{\partial}{\partial t}\mathbf{H}$ for the third term and the Ampere-Maxwell law in the absence of free electric current $\nabla\times\mathbf{H}=\frac{\partial}{\partial t}\mathbf{D}=\varepsilon\frac{\partial}{\partial t}\mathbf{E}$ for the fourth term. In the third equality, we have applied the product rule of differentiation for the third and fourth terms. The first and fourth and the second and third terms cancel each other. Thus, we have shown that the four-divergence of the first row vector of the MP SEM tensor is zero.

The four-divergences of the remaining row vectors of the MP SEM tensor in the laboratory frame can be written together as
\begin{align}
 &\frac{1}{c}\frac{\partial}{\partial t}\Big[\frac{n^2}{c}\mathbf{E}\times\mathbf{H}\Big]+\nabla\cdot(-\boldsymbol{\sigma})\nonumber\\
&=\frac{n^2}{c^2}\frac{\partial\mathbf{E}}{\partial t}\times\mathbf{H}+\frac{n^2}{c^2}\mathbf{E}\times\frac{\partial\mathbf{H}}{\partial t}
-\varepsilon(\nabla\cdot\mathbf{E})\mathbf{E}-\varepsilon(\mathbf{E}\cdot\nabla)\mathbf{E}\nonumber\\
 &\hspace{0.4cm}+\frac{1}{2}\varepsilon\nabla\mathbf{E}^2-\mu(\nabla\cdot\mathbf{H})\mathbf{H}-\mu(\mathbf{H}\cdot\nabla)\mathbf{H}+\frac{1}{2}\mu\nabla\mathbf{H}^2\nonumber\\
&=\frac{n^2}{c^2}\frac{\partial\mathbf{E}}{\partial t}\times\mathbf{H}+\frac{n^2}{c^2}\mathbf{E}\times\frac{\partial\mathbf{H}}{\partial t}
-\varepsilon\mu\mathbf{E}\times\frac{\partial\mathbf{H}}{\partial t}
-\varepsilon\mu\frac{\partial\mathbf{E}}{\partial t}\times\mathbf{H}\nonumber\\
&=\mathbf{0}.
 \label{eq:apx2}
\end{align}
In the first equality, we have taken the constant factors outside the derivatives and applied the product rule of differentiation for the cross product and outer product terms. In the second equality, we have used Gauss's laws $\nabla\cdot\mathbf{D}=\varepsilon\nabla\cdot\mathbf{E}=0$ for the third term and $\nabla\cdot\mathbf{B}=\mu\nabla\cdot\mathbf{H}=0$ for the sixth term, and the mathematical identity $\frac{1}{2}\nabla\mathbf{A}^2=\mathbf{A}\times(\nabla\times\mathbf{A})+(\mathbf{A}\cdot\nabla)\mathbf{A}$ for the fifth and eighth terms with the Faraday and Ampere-Maxwell laws. This has also canceled the fourth and seventh terms. In the third equality, we have used the identity $\varepsilon\mu=n^2/c^2$ for the refractive index to obtain the final result, where the first and fourth terms and the second and third terms have canceled each other. Thus, we have shown that the four-divergences of the MP SEM tensor row vectors below the first row vector are zero.

Equations \eqref{eq:apx1} and \eqref{eq:apx2} together indicate that the four-divergence of the MP SEM tensor is zero. The zero four-divergence in an arbitrary inertial frame is a direct consequence of the zero four-divergence in the laboratory frame since the MP SEM tensor transforms according to the Lorentz transformation as described in detail in Ref.~\cite{Partanen2019a}.

\end{document}